# Conduction and valence band offsets of Ga$_2$O$_3$/$h$-BN heterojunction

Kuang-Hui Li,[1,*] Carlos G. Torres-Castanedo,[1,*] Suresh Sundaram,[2] Haiding Sun,[1] Laurentiu Braic,[3] Mohamed N. Hedhili,[3] Abdallah Ougazzaden,[2,4] Xiaohang Li[1,±]

*Equal contribution

[1]*King Abdullah University of Science and Technology (KAUST),*

*Advanced Semiconductor Laboratory, Thuwal 23955-6900, Saudi Arabia*

[2]*Georgia Tech Lorraine, UMI 2958, Georgia Tech-CNRS, 57070 Metz, France*

[3]*King Abdullah University of Science and Technology (KAUST),*

*Core Labs, Thuwal 23955-6900, Saudi Arabia*

[4] *School of Electrical and Computer Engineering, Georgia Institute of Technology, GT-Lorraine, UMI 2958 Georgia Tech-CNRS, Metz 57070, France.*

[±]Corresponding author: xiaohang.li@kaust.edu.sa

**Abstract**

$h$-BN and Ga$_2$O$_3$ are two promising semiconductor materials. However, the band alignment of the Ga$_2$O$_3$/$h$-BN heterojunction has not been identified, hindering device development. In this study, the heterojunction was prepared by metalorganic chemical vapor deposition and pulsed laser deposition. Transmission electron microscopy confirmed sharp heterointerface and revealed structural evolution as amorphous-Ga$_2$O$_3$ grew thicker on lattice mismatched $h$-BN. The valence and conduction band offsets were determined by high-resolution X-ray photoemission spectroscopy to be 1.75 ± 0.10 and 3.35-3.65 ± 0.10 eV, respectively, corresponding to a type-II heterojunction. The extremely large type-II band offsets along with indirect bandgap of Ga$_2$O$_3$ may be leveraged for exceptional electron confinement and storage.

Monoclinic gallium oxide (Ga$_2$O$_3$) has a high theoretical breakdown field of 8 MV/cm which is superior to those of GaN (3.8 MV/cm) and SiC (3 MV/cm).[1,2] Recently, a vertical diode of 1 kV reverse breakdown voltage was achieved using Ga$_2$O$_3$.[3] In addition, due to its wide band gap (~4.90 eV), the Ga$_2$O$_3$ solar-blind deep ultraviolet photodetectors are promising.[4] Besides, the high thermal and chemical stability allow sensors based on Ga$_2$O$_3$ to operate at high temperatures (1000 °C).[5] To grow Ga$_2$O$_3$, different techniques have been used such as pulsed laser deposition (PLD),[6] molecular beam epitaxy (MBE),[7] and metalorganic chemical vapor deposition

(MOCVD),[8] and mist chemical vapor deposition (CVD).[9] Currently, sapphire and $Ga_2O_3$ native substrates are the most commonly studied substrates for growing $Ga_2O_3$ thin films. However, neither sapphire nor $Ga_2O_3$ substrate is a good thermal conductor, which hinder the application for $Ga_2O_3$–based power device on sapphire/$Ga_2O_3$ substrate due to heat dissipation issue.[10] Hexagonal boron nitride (*h*-BN) with a wide band gap (~6.00 eV) has a very high thermal conductivity of ~300 $Wm^{-1}K^{-1}$ at room temperature, close to common thermal conductors such as silver (476 $Wm^{-1}K^{-1}$) or copper (402 $Wm^{-1}K^{-1}$).[11] Due to its hexagonal structure, it has been extensively used to grow high quality graphene and has emerged as a building block for van der Waals heterojunctions.[12] *h*-BN has been used in various device applications, such as BN/graphene/BN field-effect transistors[13] and UV lasers achieved by electron-beam excitation source.[14] Furthermore, one boron isotope, $^{10}B$, has one of the largest neutron capturing cross section (3835 barn) among all the chemical elements.[15] Therefore, a compact solid-state neutron detector made by *h*-BN has the potential to replace bulk conventional $^3$He netron detector in the future. MOCVD was recently used to grow wafer-size *h*-BN films.[16] Recently, the $Ga_2O_3$/*h*-BN-based MISFET[17] and MISHEMT[18] using *h*-BN as gate dielectric layer have been reported. To design and operate devices, the conduction and valence band offsets are very important parameters since they determine the energy barriers for electron and hole transport amid a heterojunction. To date, however, there has not been any work reporting on the band offsets of the $Ga_2O_3$/*h*-BN heterojunction.

In this study, we investigated the band offsets of the $Ga_2O_3$/*h*-BN heterojunction. The growth of $Ga_2O_3$ and *h*-BN thin films was conducted by PLD and MOCVD, respectively. The $Ga_2O_3$/*h*-BN heterointerface was studied by high-resolution transmission electron microscopy (HR-TEM). The binding energies and core levels were investigated by high resolution X-ray photoelectron spectroscopy (HR-XPS). Then, the valence and conduction band offsets ($\Delta E_V$ and $\Delta E_C$) were determined.

Three thin film samples were prepared for the HR-XPS measurements: Sample 1 (35 nm *h*-BN on a *c*-plane sapphire substrate), Sample 2 (380 nm $Ga_2O_3$ on a 35 nm *h*-BN/*c*-plane sapphire template), and Sample 3 (2 nm $Ga_2O_3$ on a 35 nm *h*-BN/*c*-plane sapphire template). The *h*-BN film was grown on sapphire by an AIXTRON showerhead MOCVD system and the detailed growth condition can be found elsewhere.[Error! Bookmark not defined.] Subsequent to the growth, the *h*-BN film samples were sealed in sample containers in the $N_2$ glovebox of the MOCVD system. Then, one

*h*-BN sample was taken out of the N$_2$ glovebox and transferred to the PLD chamber which was not connected to the MOCVD system without any ex-situ cleaning process. In this process, the *h*-BN sample was exposed in the air for a few seconds (opened sample container) before loading into the PLD chamber. The Ga$_2$O$_3$ films were deposited by a Neocera Pionner 180 PLD system, equipped with a Coherent 205F laser working at 248 nm and a base pressure of less than $1\times10^{-7}$ Torr. These films were grown at 575 °C with 5 mTorr O$_2$ partial pressure. A one-inch Ga$_2$O$_3$ target (PVD Products) was ablated at 5 Hz and energy density of ~2 J/cm$^2$. The target to substrate distance was 10 cm. The target and substrate rotation speeds were 90 and 20 RPM, respectively. Before Ga$_2$O$_3$ deposition, the target was ablated for 2,000 pulses to clean the target surface. The number of laser pulses for Sample 2 was 30,000 pulses and for Sample 3 was 160 pulses. Except the number of laser pulses, Sample 2 and Sample 3 had the same PLD condition. The Ga$_2$O$_3$ and Ga$_2$O$_3$/*h*-BN samples were loaded into the XPS chamber after Ga$_2$O$_3$ deposition without any ex-situ cleaning process. Another *h*-BN sample was taken out of the N$_2$ glovebox of the MOCVD system and transferred to the XPS chamber without any ex-situ cleaning process. It was the standard process that in-situ Ar$^+$ ion beam cleaning was applied to remove any ultra-thin oxide layer and contamination before the HR-XPS measurement.

The crystal structure was examined amid 2θ-ω scans by a Bruker D8 Advance X-Ray diffractometer (XRD) equipped with K$_α$ emission line of Cu (λ=1.5405 Å). **Error! Reference source not found.** shows the 2θ-ω XRD patterns for Sample 1 and Sample 2, together with the position of the observed peaks for *h*-BN (PDF 04-015-0444) and Ga$_2$O$_3$ (PDF 04-007-0499). In Sample 1, *h*-BN (002) and sapphire (006) peaks were identified at 26.3° and 41.7°, respectively. For Sample 2, the sapphire (006) and *h*-BN (002) peak intensities are weaker due to the deposition of the thick Ga$_2$O$_3$ film (380 nm). The peaks for this film were found to be (-201), (-402), and (-603) planes, which are the most common peaks for monoclinic Ga$_2$O$_3$ deposited on *c*-plane sapphire. Additional peaks such as (-401) and (510) planes are present in Sample 2, belonging also to monoclinic phase. Besides large lattice mismatch between *h*-BN and Ga$_2$O$_3$, these two planes may be caused by non-optimized O$_2$ partial during the deposition. Many studies have shown that O$_2$ partial pressure has impact on oxygen vacancies, conductivity, growth rate, Ga/O ratio, and single/poly crystallinity.[19] For Sample 3, the same diffraction pattern as Sample 1 was not observed since the Ga$_2$O$_3$ film is merely 2 nm thick.

The surface morphology of Sample 1 and Sample 2 was studied by a Dimension Icon atomic force microscope (AFM) under atmospheric conditions, as shown in **Error! Reference source not found.**(a) and **Error! Reference source not found.**(b), respectively. The RMS roughness decreases from 4.03 nm for Sample 1 to 3.00 nm for Sample 2 with the 380 nm $Ga_2O_3$ deposition. As observed in the images, the grain size of $Ga_2O_3$ is smaller than *h*-BN, reducing the roughness in Sample 1. From **Error! Reference source not found.**(a), the morphology shows wrinkles, which might be one of the reason causing Sample 2 to have different orientations than (-201), i.e. the (-401) and (510) peaks in **Error! Reference source not found.**. The wrinkle can be explained by the accepted point-of-view of *h*-BN growers that it originates from thermal stresses while cooling down. From **Error! Reference source not found.**(b), the morphology shows no wrinkles on the surface which was covered by the $Ga_2O_3$ film. Scanning transmission electron microscopy (STEM) was utilized to investigate Sample 2 in particular the heterointerface by operating a probe corrected FEI Titan system at an acceleration voltage of 300 kV. The TEM specimens were prepared by an FEI Helios G4 dual beam focused ion beam (FIB) equipped with an omniprobe. Electron energy loss spectroscopy (EELS) acquisition was performed to check the interface sharpness by detecting the elemental distribution at the $Ga_2O_3$/*h*-BN/sapphire interfaces under STEM mode. **Error! Reference source not found.**(a) displays the cross-sectional STEM image of Sample 2. The thickness of $Ga_2O_3$ and *h*-BN were confirmed to be 380 and 35 nm, respectively. **Error! Reference source not found.**(b) presents the elemental distribution from the EELS scan of Ga, B, Al, N, and O across the $Ga_2O_3$/*h*-BN/sapphire interfaces. It confirms a uniform distribution of the elements in each layer with sharp interface, indicating relatively low concentration of impurities in both layers and at the $Ga_2O_3$/*h*-BN interface. The density of O in sapphire is higher than that of $Ga_2O_3$ caused by the relatively lower density of the $Ga_2O_3$ film compared with the high density single crystalline sapphire substrate.

In **Error! Reference source not found.**(c), the $Ga_2O_3$/*h*-BN interface image was taken by STEM from the green-dash square in **Error! Reference source not found.**(a) based on Sample 2 since the $Ga_2O_3$ layer thickness of Sample 3 was too thin. *h*-BN has layer structure parallel to *c*-plane, which is an iconic 2D material characteristic. There is a thin blur layer between *h*-BN and $Ga_2O_3$. The thin blur layer was caused by *h*-BN surface fluctuation: because of the weak Van der Waals bond between *h*-BN layers, some *h*-BN domains inevitably self-delaminated, i.e. surface fluctuation, as reflected by the imperfect alignment of *h*-BN lattice. The fluctuation can cause a seemingly blurred interface layer for a STEM image since the TEM laminar was about 100 nm thick which may contain more than one domain with surface fluctuation. However, it is important to note that mild delamination does not impact the band alignment result in terms of lattice orientation because the h-BN/$Ga_2O_3$ interface remains *c*-plane regardless.

In **Error! Reference source not found.**(c), we found that $Ga_2O_3$ is amorphous right at the interface, mainly due to large lattice mismatch between $Ga_2O_3$ and *h*-BN. According to the references[20,21], the B-N bond length in *h*-BN is 1.45 Å and the Ga-O bond length in b-Ga2O3 is 1.83 Å, so that the mismatch is 26.2%. Hence, the result of the band offset measurement by XPS in this study should be referred to as amorphous $Ga_2O_3$/*h*-BN band offset instead of single crystal *β*-$Ga_2O_3$/*h*-BN band offset. Moreover, the finding indicates that the amorphoousness may also exist at the interface of other $Ga_2O_3$-based heterojunctions due to lattice mismatch, albeit most band offset studies of $Ga_2O_3$-based heterojunctions do not include the TEM study of the interface.[22-26] Moreover, as the $Ga_2O_3$ film thickness increases, two regions with orientated planes (encircled by white dash line) starts to appear. These planes are almost perpendicular to the $Ga_2O_3$/h-BN interface and have d-spacing 0.30 nm and 0.18 nm, respectively, which correspond to the d-spacing of (-401) and (510) planes based on Equation (1), also shown in the XRD spectrum of Figure 1. At larger $Ga_2O_3$ film thickness, (-201) starts to dominate as measured by the d-spacing as shown in Figure 3(e) based on Equation (1). The evolution of crystal structure indicates that the growth condition still favors the formation of the orthodox (-201) plane but the large lattice mismatch at the interface causes the initial $Ga_2O_3$ layers to be amorphous on *h*-BN, which should apply to not only PLD but also other techniques such as MOCVD and MBE.[27,28] The analysis based on Figure 3(c) and (d) applies to Sample 3 as well since Sample 2 and Sample 3 had the same growth condition despite the difference in thickness. **Error! Reference source not found.**(d) shows the atomic resolution STEM image from red-dash square in **Error! Reference source not found.**(a), which shows different grain boundaries especially at bottom-right corner. The monoclinic structure of $Ga_2O_3$ has the following lattice parameters: $a$=12.23 Å, $b$=3.04 Å, $c$=5.80 Å, and $β$=103.7° (angle between "*a*" and "*c*" axes).[29] By inserting these parameters into Eq. (1) for a monoclinic lattice, we obtain the spacing of (-201) which is equal to 4.6 Å, matching the spacing measured by HR-STEM (**Error! Reference source not found.**(e)).

$$d = \left(\frac{h^2}{a^2 \sin^2 \beta} + \frac{k^2}{b^2} + \frac{l^2}{c^2 \sin^2 \beta} - \frac{2hl \cos \beta}{ac \sin^2 \beta}\right)^{-\frac{1}{2}} \quad (1)$$

The HR-XPS experiment was carried out by a Kratos Axis Supra DLD spectrometer with an Al K$\alpha$ X-ray source ($h\upsilon$=1486.6 eV) with 150 W power, base pressure of $10^{-9}$ mbar, and a multiple channel plate. The measured binding energies were referenced to the C 1s binding energy of carbon contamination (284.8 eV). The resolution of the peak position in the XPS spectra was estimated to be 0.10 eV. For the core level spectra, experimental data points were fitted by the Voigt (Gaussian-Lorentzian function) curve after considering a Shirley background, while the valence band maxima (VBM) energy in the spectrum was determined by extrapolating a linear fit of the leading edge of the valance band photoemission to the baseline.

According to the Kraut's method, the valence band offset ($\Delta E_V$) and conduction band offset ($\Delta E_C$) of the Ga$_2$O$_3$/h-BN heterojunction can be calculated as follows:[30]

$$\Delta E_V = \left(E_{Ga\,2p_{3/2}}^{Ga_2O_3} - E_{VBM}^{Ga_2O_3}\right) - \left(E_{B\,1s}^{h-BN} - E_{VBM}^{h-BN}\right) - \Delta E_{CL} \quad (2)$$

$$\Delta E_C = \left(E_g^{h-BN} - E_g^{Ga_2O_3}\right) - \Delta E_V \quad (3)$$

where

$$\Delta E_{CL} = \left(E_{Ga\,2p_{3/2}}^{Ga_2O_3/h-BN} - E_{B\,1s}^{Ga_2O_3/h-BN}\right) \quad (4)$$

$\left(E_{B\,1s}^{h-BN} - E_{VBM}^{h-BN}\right)$ and $\left(E_{Ga\,2p_{3/2}}^{Ga_2O_3} - E_{VBM}^{Ga_2O_3}\right)$ in eq(2) are the VBM binding energies with reference to the core levels (CLs) in Sample 1 (B 1s) and Sample 2 (Ga 2p$_{3/2}$), respectively. **Error! Reference source not found.**(a) and **Error! Reference source not found.**(b) show the VBM (2.31 eV) and the B 1s binding energy (190.58 eV) for Sample 1. The defect structure in Ga$_2$O$_3$ has several types, with oxygen vacancies being the most common one.[31] So far, few studies have been reported regarding how those defects in III-oxides impact the band offset due to highly complex nature of the defects. Oxygen vacancies form deep levels in the band gap causing visible photoluminescence or form donor levels near conduction band causing un-doped condcutive Ga$_2$O$_3$. The density of states (DOS) close to VBM in Ga$_2$O$_3$ is mainly contributed to O 2p orbitals.[32,33] In PLD-deposited Ga$_2$O$_3$ with oxygen vacancies, these imperfections might broaden DOS distribution by chance, which might be different from high-quality single crystalline $\beta$-Ga$_2$O$_3$. **Error! Reference source not found.**(c) and **Error! Reference source not found.**(d) display the obtained values for the VBM (3.34 eV) and the binding energy of Ga 2p$_{3/2}$ (1117.64 eV). The last term in Eq. (4) ($\Delta E_{CL}$) is the binding energy difference between $E_{Ga\ 2p_{3/2}}^{Ga_2O_3/h-BN}$ (Ga $2p_{3/2}$) and $E_{B\ 1s}^{Ga_2O_3/h-BN}$ (B 1s) CLs, measured at the Ga$_2$O$_3$/h-BN interface of Sample 3. **Error! Reference source not found.**(e) and **Error! Reference source not found.**(f) show the Ga $2p_{3/2}$ (1117.76 eV) and B 1 s CLs (189.98 eV), respectively. For the B 1s CL, one extra Voigt curve was needed to fit the data at 191.75 eV, assigned to oxidized B complex during the formation of the Ga$_2$O$_3$/h-BN interface. From Eq. (2), the $\Delta E_V$ is determined to be -1.75 ± 0.10 eV.

The $\Delta E_V$ determined by the XPS measurement is certain and definite. To obtain the conduction band offset $\Delta E_C$, the band gaps of h-BN and Ga$_2$O$_3$ are needed. Although there are multiple reported h-BN band gaps (5.50-6.00 eV),[34] the most accepted value (6.00 eV) for the indirect band gap of h-BN is used. For Ga$_2$O$_3$, Kumar and Zhang et al. have reported the band gaps of amorphous Ga$_2$O$_3$, which range from 4.1 to 4.4 eV.[35,36] By inserting $\Delta E_v$ and the band gaps of both Ga$_2$O$_3$ (4.1-4.4 eV) and h-BN[37] (6.00 eV) in Eq. (3), we can obtain the $\Delta E_C$ of 3.65-3.35 eV for the Ga$_2$O$_3$/h-BN heterojunction, which is very large among semiconductor heterojunctions. The results of this study were incorporated into the band diagram shown in Figure 5. It suggests that the Ga$_2$O$_3$/h-BN heterojunction is type-II (staggered gap). The reported band alignments for wurtzite AlN and GaN with Ga$_2$O$_3$[6,38] were also included in **Error! Reference source not found.**5 as reference. The $\Delta E_C$ and $\Delta E_V$ of the Ga$_2$O$_3$/h-BN heterojunction are larger compared to those of heterojnctions formed between AlN and GaN, and Ga$_2$O$_3$. There could be different applications of the investigated heterojunction in optoelectronics and electronics. In particular, it is worth mentioning that the very large band offsets and indirect band gap of Ga$_2$O$_3$ could make an h-BN/Ga$_2$O$_3$/h-BN double heterojunction an exceptional candidate to confine and store electrons within the Ga$_2$O$_3$ layer (electron well), which could be employed in devices such as non-volatile memory.[39,40] It is important to note that this double heterojunction cannot be realized by bulk Ga$_2$O$_3$ substrates since Ga$_2$O$_3$ is sanwidtched between two h-BN layers. Thus, epitaxial and deposition techniques have to be used including but not limited to MOCVD, MBE, mist CVD, and PLD.

In summary, we have deposited Ga$_2$O$_3$ film by PLD on MOCVD-grown h-BN/sapphire templates to form the Ga$_2$O$_3$/h-BN heterojunction. EELS elemental mapping confirmed sharp heterointerface interface and HR-TEM experiments revealed that initial Ga$_2$O$_3$ layers at the interface were amorphous due to large lattice mismatch. By employing HR-XPS, the valence and conduction band offsets of the $Ga_2O_3$/*h*-BN heterojunction were determined to be 1.75 and 3.35-3.65 eV, respectively, corresponding to a type-II heterojunction. The very large type-II band offsets together with indirect bandgap of $Ga_2O_3$ may be utilized for exceptional electron confinement and storage.

## Acknowledgments

Members of the Advanced Semiconductor Laboratory at KAUST would like to acknowledge the support of KAUST Baseline BAS/1/1664-01-01, KAUST Competitive Research Grant URF/1/3437-01-01, and GCC Research Council REP/1/3189-01-01. The authors SS and AO acknowledge funding by the French National Research Agency (ANR) under the GANEX Laboratory of Excellence (Labex) project.

**Figure Captions**

**Fig. 1.** XRD 2θ-ω spectra of Sample 1 and Sample 2.

**Fig. 2.** AFM images of (a) Sample 1 (*h*-BN, 2.5×2.5 µm²) and (b) Sample 2 (Ga₂O₃/*h*-BN, 5×5 µm²).

**Fig. 3**. (a) A cross-sectional TEM image of Sample 2. (b) An EELS elemental map of Ga, B, Al, N, and O from the white-solid square in (a). (c) A STEM image at h-BN/β-Ga₂O₃ interface from the green-dash square in (a). (d) The high-resolution image from the red-dash square in (a). (e) A zoom-in image at from the red-dash square in (d).

**Fig. 4.** (a) Valence band spectra and (b) CL B 1s of Sample 1. (c) Valence band spectra and (d) CL Ga $2p_{3/2}$ of Sample 2. (e) CL B 1s and (f) CL Ga $2p_{3/2}$ of Sample 3. The CLs were fitted by Voigt curves using a Shirley background.

**Fig. 5** Band alignment diagram of the Ga₂O₃/*h*-BN heterojunction, along with w-GaN and w-AlN. (a) from Ref. 29, (b) from Ref. 6, and (c) from this study.

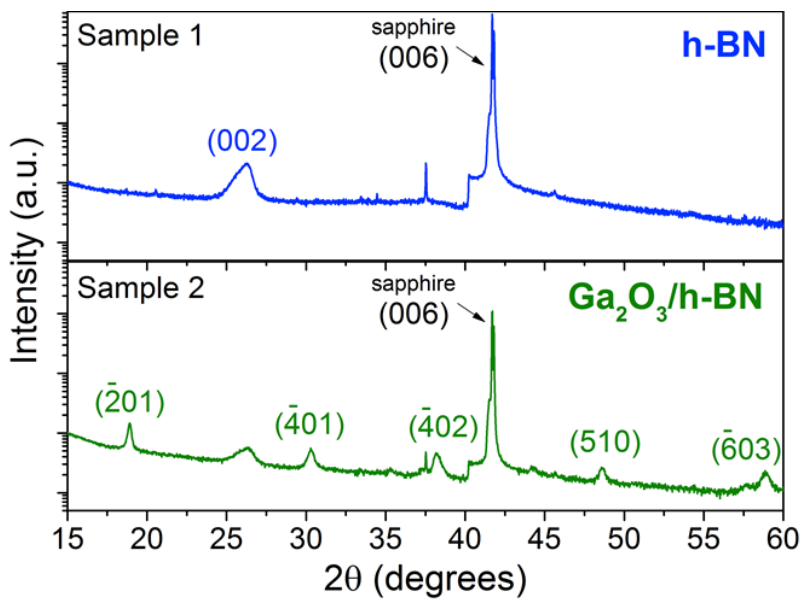

**Fig. 1.**

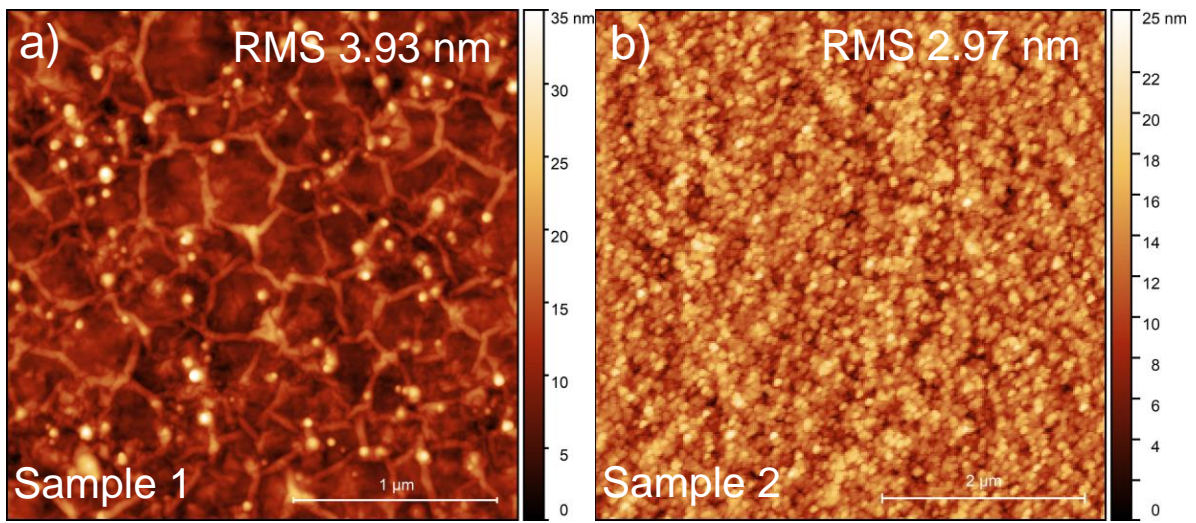

**Fig. 2.**

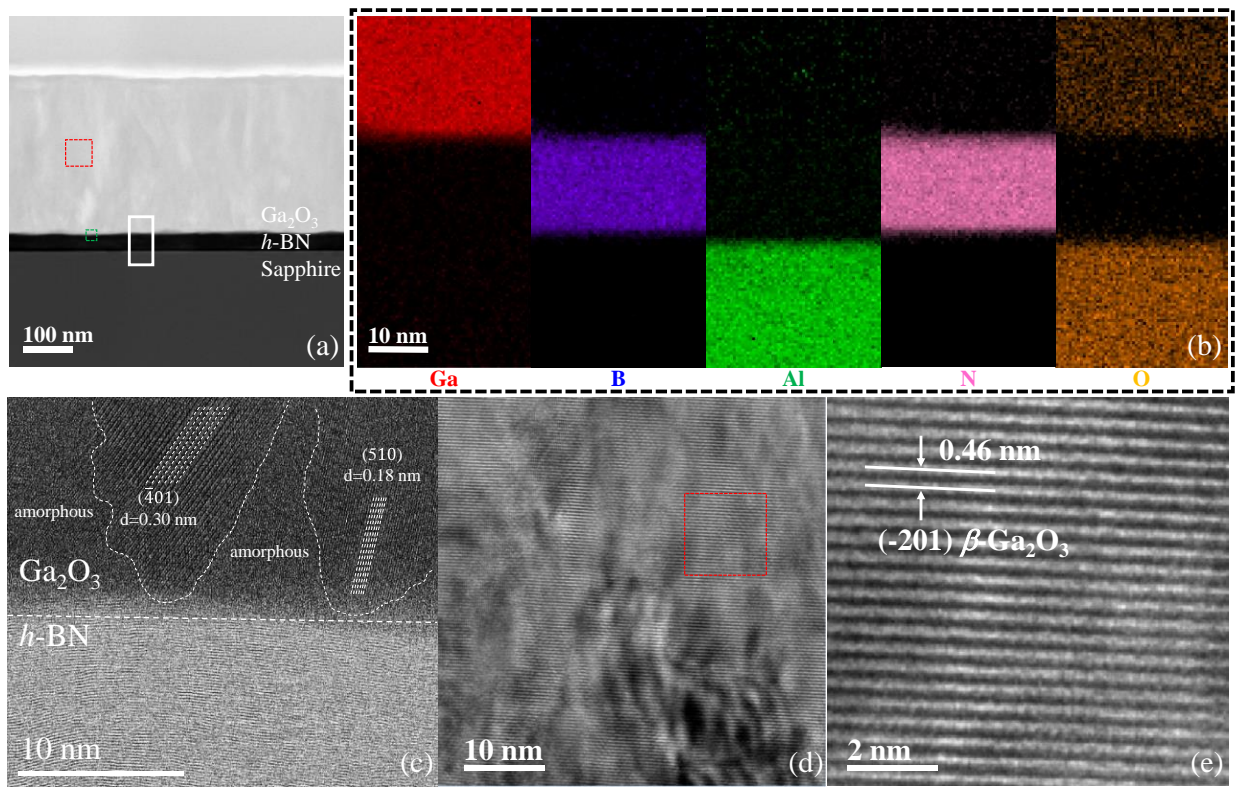

Fig. 3.

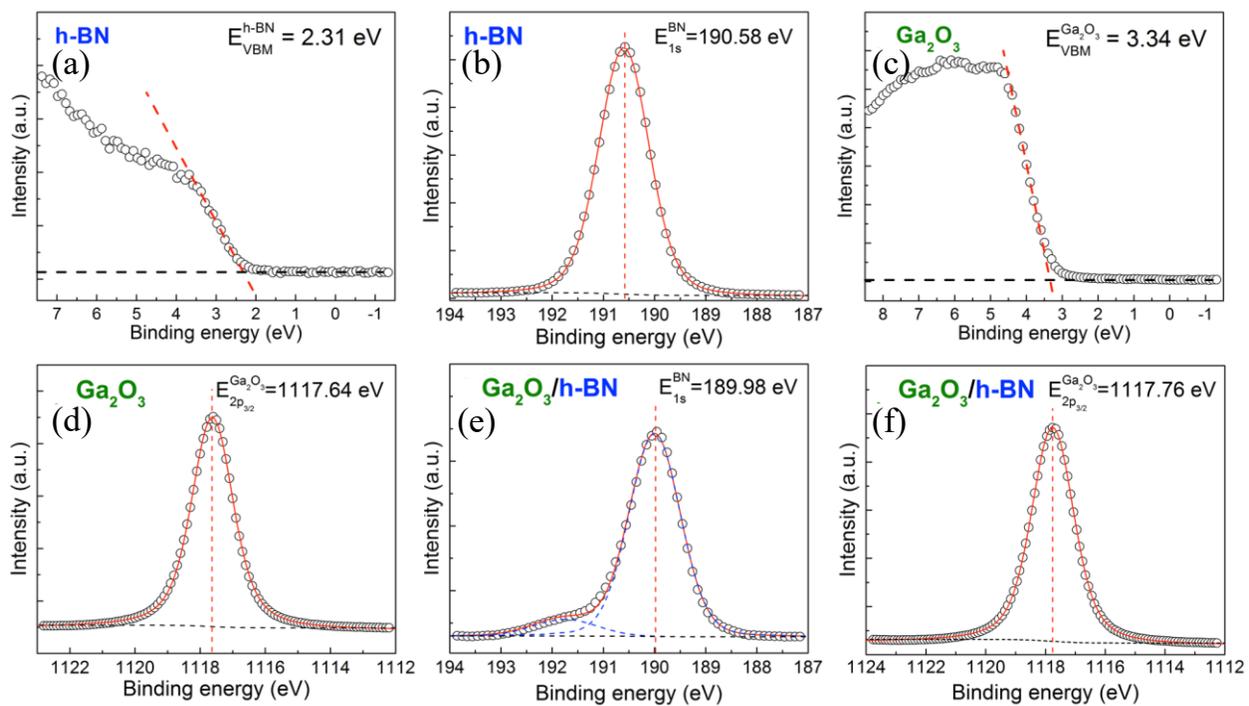

**Fig. 4.**

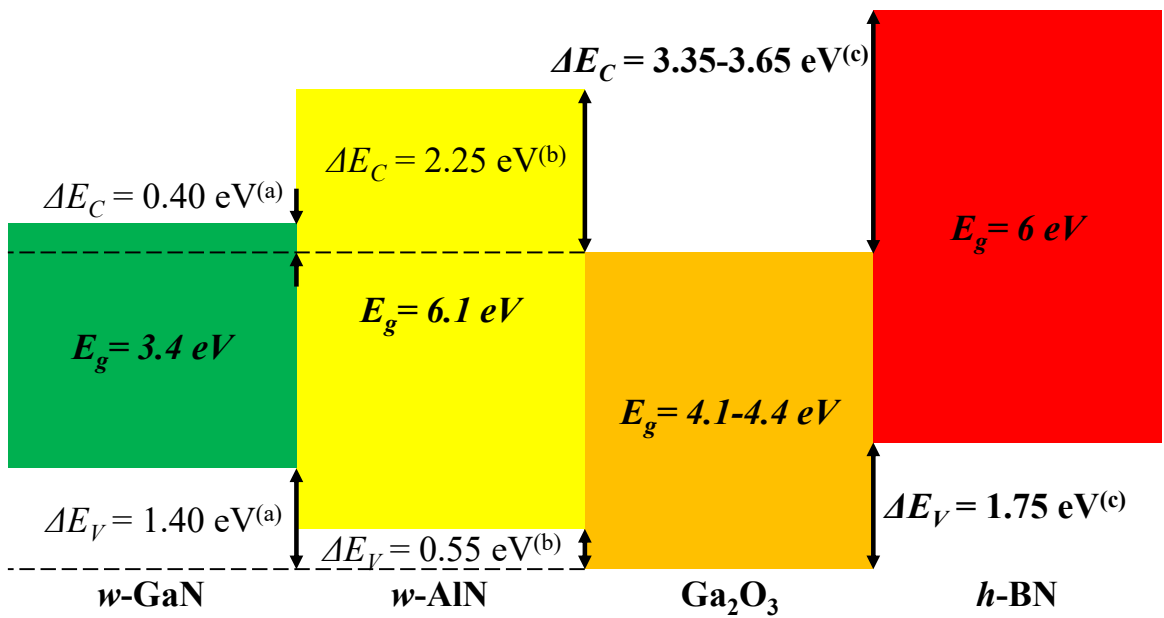

**Fig. 5.**